\pdfoutput=1

\documentclass[%
 reprint,
superscriptaddress,
 amsmath,amssymb,
 aps,
pra,
]{revtex4-1}

\usepackage{graphicx}% Include figure files
\usepackage{dcolumn}% Align table columns on decimal point
\usepackage{bm}% bold math
\usepackage[hidelinks]{hyperref}% add hypertext capabilities

\begin{document}

\preprint{APS/123-QED}

\title{Pathlengths of open channels in multiple scattering media}% Force line breaks with \\

\author{Jeroen Bosch}
\email{j.bosch1@uu.nl}
\affiliation{
Complex Photonic Systems (COPS), MESA+ Institute for Nanotechnology, University of Twente, P.O. Box 217, 7500 AE Enschede, The Netherlands
}
\affiliation{
Nanophotonics, Debye Institute for Nanomaterials Research, Center for Extreme Matter and Emergent Phenomena, Utrecht University, P.O. Box 80,000, 3508 TA Utrecht, The Netherlands
}

\author{Sebastianus A. Goorden}
\affiliation{
Complex Photonic Systems (COPS), MESA+ Institute for Nanotechnology, University of Twente, P.O. Box 217, 7500 AE Enschede, The Netherlands
}%
\affiliation{
Present adress: ASML Netherlands B.V., De Run 6501, 5504 DR Veldhoven, The Netherlands
}

\author{Allard P. Mosk}
\affiliation{
Complex Photonic Systems (COPS), MESA+ Institute for Nanotechnology, University of Twente, P.O. Box 217, 7500 AE Enschede, The Netherlands
}
\affiliation{
Nanophotonics, Debye Institute for Nanomaterials Research, Center for Extreme Matter and Emergent Phenomena, Utrecht University, P.O. Box 80,000, 3508 TA Utrecht, The Netherlands
}

\date{\today}% It is always \today, today,

\begin{abstract}
We report optical measurements of the spectral width of open transmission channels in a three-dimensional diffusive medium. The light transmission through a sample is enhanced by efficiently coupling to open transmission channels using repeated digital optical phase conjugation. The spectral properties are investigated by enhancing the transmission, fixing the incident wavefront and scanning the wavelength of the laser. 
We measure the transmitted field to extract the field correlation function and the enhancement of the total transmission. 
We find that optimizing the total transmission leads to a significant increase in the frequency width of the field correlation function.
Additionally we find that the enhanced transmission persists over an even larger frequency bandwidth.
This result shows open channels in the diffusive regime are spectrally much wider than previous measurements in the localized regime suggest.
\end{abstract}

\maketitle

%\tableofcontents

\section{Introduction}
Many well-known effects in wave transport result from interference and cannot be described by diffusion theory. These effects include enhanced backscattering \cite{Albada1985,Wolf1985}, Anderson localization \cite{Akkermans2007,Lagendijk2009} and universal conductance fluctuations \cite{Scheffold1998}. A striking interference phenomenon is the existence of highly transmitting channels in multiple scattering systems, which allow unity transmission through arbitrarily thick non-absorbing diffusive layers. These highly transmitting channels, usually called ``open channels'', were initially predicted for electrons \cite{Dorokhov1982,Dorokhov1984,Lee1985,Mello1988,Nazarov1994}, while later the theory was generalized to other waves \cite{Pendry1990,Beenakker1997,Akkermans2007}.

An exciting recent development in optics is the use of wavefront shaping to coherently control light in multiple scattering media \cite{Vellekoop2007,Popoff2010_prl,Mosk2012}. Wavefront shaping enables investigation of interference effects that are difficult to elucidate in e.g. electronic systems. Open channels were observed in optics by wavefront shaping to selectively couple light into them \cite{Vellekoop2008d,Popoff2014,Kim2012}, 
and by transmission matrix measurements in microwave and acoustical waveguides \cite{Shi2012,Gerardin2014}. Numerical simulations agree with these experiments\cite{Choi2011,Liew2014,Shi2015,Hsu2015}. Open channels greatly enhance the penetration of light into multiple scattering media and may benefit a wide range of applications in e.g. healthcare, sensing, security, photovoltaics and lighting \cite{Wang2012_science,Wang2012_ncomm,Yizhar2011,Goorden2014_qsa,Burresi2015,Leung2014}.

The spectral width of open channels is of critical importance to applications. As open channels arise from interference, one may expect them to be narrowband. A recent microwave experiment in quasi-1D geometry shows that in the Anderson localized regime, where transport is dominated by a single quasimode \cite{Pena2014}, the open transmission channels are spectrally narrower than the average channel width, inhibiting applications \cite{Shi2015}.

\begin{figure}[h]
\centerline{\includegraphics[width=9cm]{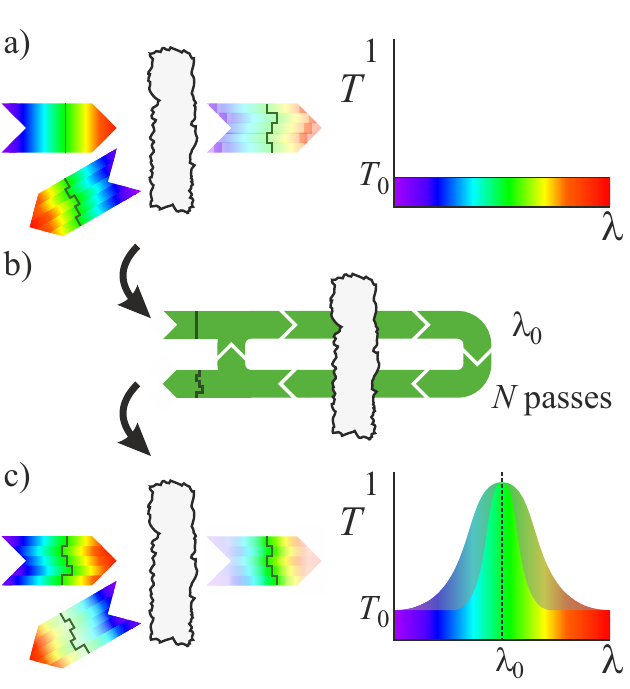}}
\caption{a) Average transmission spectrum of light through a multiple scattering medium. b) Light is coupled to open transmission channels at wavelength $\lambda_0$ by phase conjugating the transmitted light field $N$ times. c) Transmission spectrum when light is coupled into an open channel. The spectral width of open transmission channels is investigated.}
\label{fig:cartoon}
\end{figure}

Here, we use transmission enhancement as a robust way to measure on open transmission channels in 3D diffusive media.
Our experiment is based on repeated phase conjugation, which is a physical implementation of the Von Mises iteration \cite{Mises1929}, as shown in Fig. \ref{fig:cartoon}. In a single pass through the medium the average transmission is not wavelength dependent (Fig. \ref{fig:cartoon} (a)). Repeated phase conjugation of light through the sample leads to efficient enhancement of the transmission at wavelength $\lambda_0$ \cite{Hao2014}, since the most transmitting channels are more strongly represented in every step (Fig. \ref{fig:cartoon} (b)). After enhancing the transmission, we fix the spatial wavefront of the incident light and tune the wavelength. The spectral width of the transmission enhancement is determined by measuring the transmission spectrum, as well as the field correlation function, as illustrated in Fig. \ref{fig:cartoon}(c). Surprisingly, we find that in our 3D diffusive samples the transmission enhancement is spectrally broader than the well known $C^1$ speckle correlation function \cite{Feng1988} that represents the channel average. 

\section{Experimental apparatus}
\begin{figure}[h]
\centerline{\includegraphics[width=9cm]{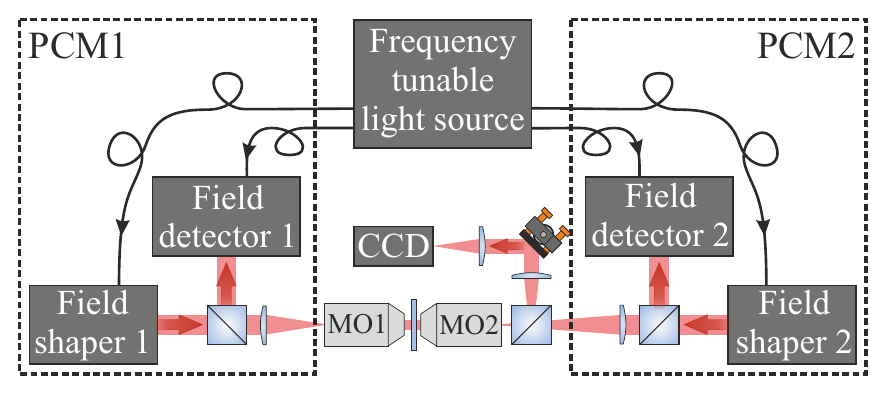}}
\caption{Schematic overview of the iterative phase conjugation apparatus. Each side of a slab-geometry sample is imaged to a phase conjugate mirror (PCM), which allows detection and reconstruction of the full vector light field.}
\label{fig:apparatus_overview}
\end{figure}

An overview of the apparatus is shown in Fig. \ref{fig:apparatus_overview}. It consists of two digital phase conjugate mirrors, PCM1 and PCM2. Each phase conjugate mirror consists of a field detector and a field shaper. The field detectors use off-axis holography with two orthogonally polarised reference beams to retrieve the vector light field $\vec{E}(x,y)$ from a single camera image \cite{Takeda1982,Colomb2002}. The field shapers use a digital micromirror device (DMD, Vialux V-9600) and 
Lee holography \cite{Lee1974} to shape the vector light field. 
PCM1 and PCM2 are imaged to the sample surfaces with a calculated magnification of 286x, using $f=750$\,mm tube lenses and 1.4\,NA 63x (MO1) and 0.95\,NA 63x (MO2) microscope objectives, respectively. The back aperture of MO2 is imaged onto a separate charge-coupled device (CCD). 

The light source is a New Focus Velocity TLB-6712 frequency tunable diode laser. The laser scanning range is 765\,-\,781\,nm, the base resolution is 0.01\,nm and the linewidth is around 1\,MHz. Single mode polarisation-maintaining optical fibers guide the light to the field shapers and detectors. We use monochromatic cameras  with $1392\times 1040$ pixels of $6.45\,\mu$m by $6.45\,\mu$m size (Dolphin F145-B) for the field detectors.

The sample consists of a $20 \pm 10$\,$\mu$m thick layer of zinc oxide (ZnO) nanoparticles with a transport mean free path of $l_{\mathrm{tr}}=0.73 \pm 0.15\,\mu$m and is prepared as described in \cite{Puttenphdthesis}.

For details on the apparatus see \cite{Goordenphdthesis}.

\section{\label{sec:openchannels}Accurate coupling to open channels}
Iterative phase conjugation is employed to efficiently enhance transmission by coupling light to open channels of the multiple scattering sample. We obtained the best reproducibility by choosing to 
control a single polarisation component at fixed amplitude.
The optimized wavelength is set to $\lambda_0=769$\,nm. The iterative phase conjugation procedure is initialized by sending a random speckle pattern, constructed by field shaper 1, through the sample. The transmitted vector field is measured by field detector 2 and the total transmitted intensity is measured by the CCD. Then, field shaper 1 is turned off and field shaper 2 constructs the phase conjugate of the detected field, which propagates back through the sample. This completes a single phase conjugation iteration. PCM1 and PCM2 alternatingly phase conjugate the light field until the process converges. 

\begin{figure}[ht]
\centerline{\includegraphics[width=8cm]{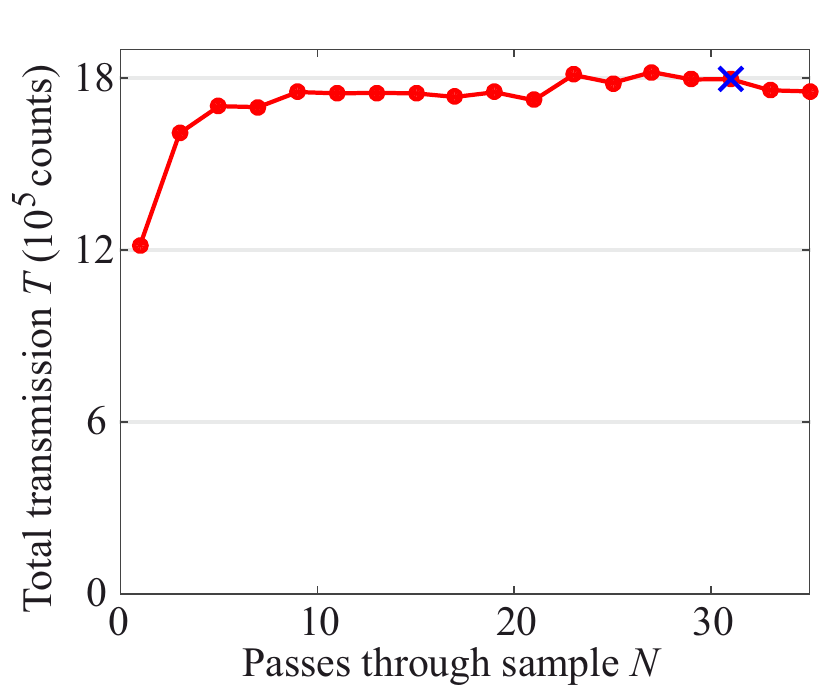}}
\caption{Measured total transmission $T$ as function of number of passes through the sample $N$. The cross indicates the field used for investigating the spectral width of open transmission channels.}
\label{fig:pingpong}
\end{figure}

The total transmitted intensity $T$ measured on the CCD during the iterative phase conjugation process is shown in Fig.~\ref{fig:pingpong}. The transmitted intensity converges to its maximum after approximately $N=9$ passes through the sample. For the investigation of the width of open channels we use the field measured on field detector 1 after $N=30$ passes, ensuring full convergence. The intensity transmission of this field is 54\% higher than the average transmission of the sample. 

\section{Correlation width of open channels}
The width of open channels is characterized in two ways. First, we define the field correlation function: 
\begin{equation}
C_{\omega_0}(\Delta\omega)=|E(\omega_0)\cdot E^{*}(\omega_0+\Delta\omega)|^2,
\label{eq:Decorr}
\end{equation}
where $E(\omega_0)$ and $E(\omega_0+\Delta\omega)$ denote the vertical polarisation component of fields measured on the field detector at $\omega_0$ and $\omega_0+\Delta\omega$, respectively. We note that this is subtly different from the well-known correlation function $C^1$ \cite{Feng1988} which by definition is averaged over $\omega_0$.
The field correlation is calculated over the square area controlled by the PCM, and the fields $E(\omega_0)$ and $E(\omega_0+\Delta\omega)$ are normalised to the transmission of a random speckle pattern through the sample. 

Secondly, we define the transmission enhancement 
\begin{equation}\eta^{\mathrm{T}}_{\omega_0} (\Delta \omega)=\frac{T^{\mathrm{opt}}(\omega_0+\Delta \omega)}{T^{\mathrm{unopt}}(\omega_0+\Delta \omega)},
\end{equation} 

where $T^{\mathrm{opt}}(\omega_0+\Delta \omega)$ denotes the total transmission measured on the CCD at $\omega_0+\Delta \omega$
for an illumination pattern that maximizes the transmission at $\omega_0$. $T^{\mathrm{unopt}}(\omega_0+\Delta \omega)$ is the average total transmission measured on the CCD at $\omega+\Delta \omega$ for an unoptimized illumination pattern with the same incident power.

\begin{figure}[h]
\centerline{\includegraphics[width=9cm]{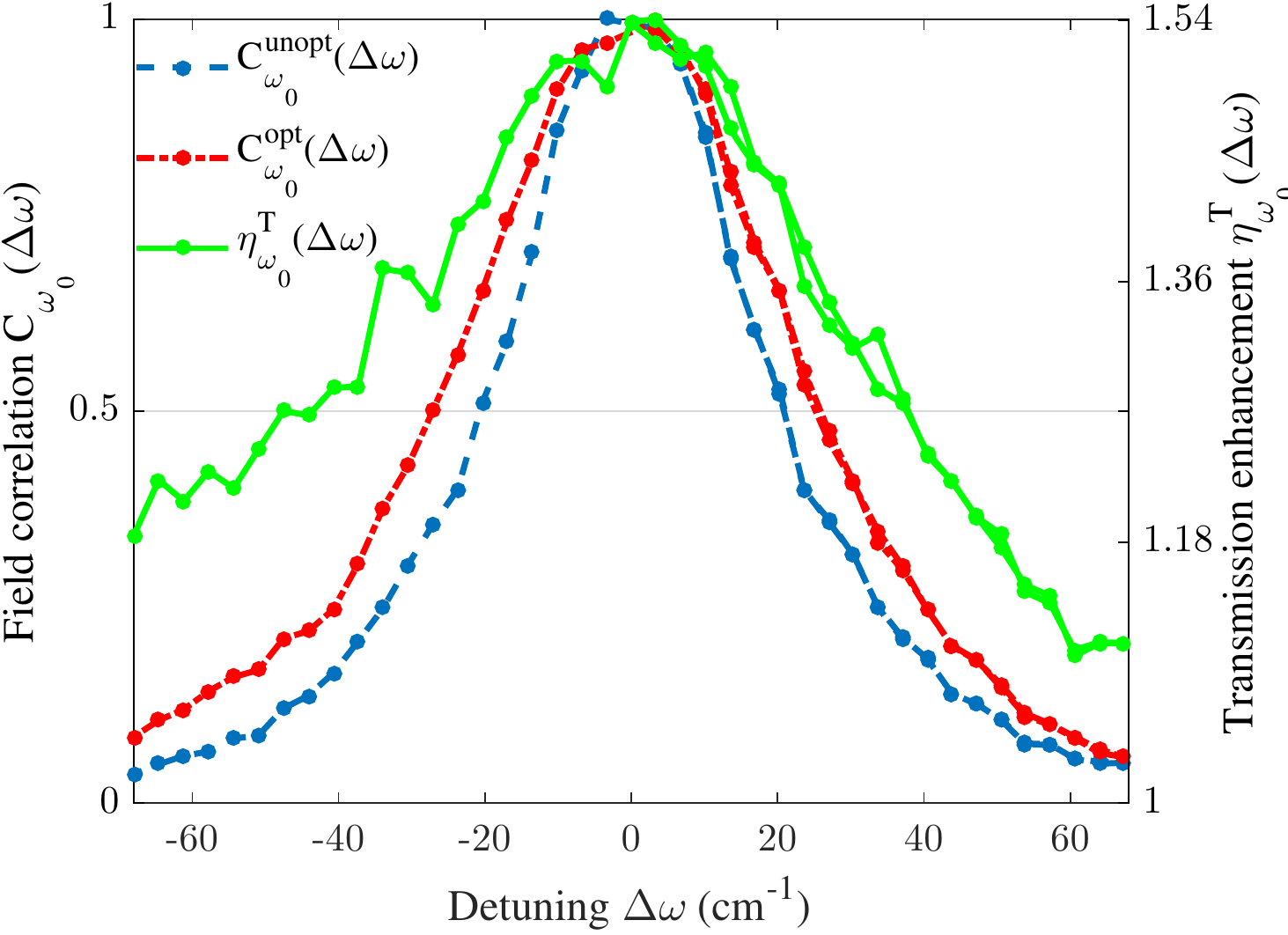}}
\caption{Transmitted field correlation for an unoptimized incident field $C^{\mathrm{unopt}}_{\omega_0}(\Delta \omega)$ and for an incident field that optimally couples to open channels $C^{\mathrm{opt}}_{\omega_0}(\Delta \omega)$. Transmission enhancement $\eta^{\mathrm{T}}_{\omega_0}(\Delta \omega)$ of an optimized incident field.}
\label{fig:decorr}
\end{figure}

The procedure for measuring the field correlation and transmission enhancement is as follows. The laser is set to $\lambda = \lambda_0 = 769$\,nm and field shaper 1 is set to optimally enhance transmission using the setting found by iterative phase conjugation. Then $E^{\mathrm{opt}}(\lambda_0)$ is measured on field detector 2. Then, the laser is scanned from the central wavelength $\lambda_0=769\,$nm up to $\lambda=773\,$nm and back down to $\lambda=765\,$nm with a stepsize of 0.2\,nm and a wavelength accuracy of 0.07\,nm. The spatial profile of the field incident on the sample is kept constant by actively correcting the DMD pattern for diffraction effects. At each wavelength step the transmitted field $E^{\mathrm{opt}}(\lambda)$ is measured on field detector 2 and the total transmitted intensity $T^{\mathrm{opt}}(\lambda)$ is measured on the CCD. 

For the reference wavelength scan, the laser is reset to $\lambda_0$ and the pattern on field shaper 1 is shifted by 20 DMD pixels in both the x and y direction (approximately 3 speckle grains) to create an effectively uncorrelated illumination pattern. $E^{\mathrm{unopt}}(\lambda_0)$ is measured on field detector 2. The same wavelength scan is performed and $E^{\mathrm{unopt}}(\lambda)$ and $T(\lambda)$ are  measured at each wavelength step. 

In Fig. \ref{fig:decorr} we show the measured field correlation function for an optimized field $C^{\mathrm{opt}}_{\omega_0}(\Delta \omega)$ and for an unoptimized field $C^{\mathrm{unopt}}_{\omega_0}(\Delta \omega)$ as well as the measured enhancement $\eta^\mathrm{T}_{\omega_0}(\Delta \omega)$. All curves decay as a function of $\Delta \omega$ with a different width.
The field correlation function of the unoptimized field $C^{\mathrm{unopt}}_{\omega_0}(\Delta\omega)  =C^1(\Delta\omega)$ has a full width at half maximum (FWHM) of $42\pm3.4\,$cm\textsuperscript{-1}.
The width of the field correlation function of the optimized field $C^{\mathrm{opt}}_{\omega_0}(\Delta \omega)$ (FWHM $52\pm3.4\,$cm\textsuperscript{-1}) is clearly larger than that of the unoptimized field $C^{\mathrm{unopt}}_{\omega_0}(\Delta \omega)$. 
The measurement of the transmission enhancement $\eta^{\mathrm{T}}(\Delta \omega)$ (FWHM $81\pm10\,$cm\textsuperscript{-1}) shows significantly more noise than the field correlation functions, but the curve is clearly broader than either of them.
Remarkably this increased width of the optimized transmission seems qualitatively different from observations in waveguides in the localised regime \cite{Shi2015}.

\section{Interpretation and outlook}
%From our measurement results we see a clear difference in the width of the field correlation function between the optimized field and an unoptimized field. Additionally we see that the spectral width of the transmission enhancement is larger than both field correlation functions. 
The initial expectation of the width of a correlation function is the Thouless frequency $\Delta \omega_{\mathrm{Th}}$, where $\Delta \omega_{\mathrm{Th}} / \omega = 6 D / (L^2 \omega )$, with the diffusion constant $D=\frac{1}{3}v_{\mathrm{E}}l_{\mathrm{tr}}$ and the transport velocity $v_{\mathrm{E}}$ \cite{Albada1991}. 
% The Thouless frequency is equal to the width of $C^1$ \cite{Feng1988}.
We find the Thouless frequency from the field correlation function for an unoptimized field.

The width of the field correlation function for the optimized field that we observe is clearly larger than that for the unoptimized field. This indicates there is a relation between transmission and time delay, suggesting that highly transmitting channels may have an effective shorter time delay. This is remarkable as in random matrix theory of chaotic systems it was found that time delay and transmission operators are statistically uncorrelated \cite{Brouwer1997,Fyodorov2011}. The observed broadening may be due to a small number of anomalous highly-transmitting simultaneous eigenstates of these operators \cite{Rotter2011}, or due to a more subtle correlation effect involving many channels.

In previous work, simulations and measurements on transmission eigenchannels for microwaves in samples in the crossover to localization have shown a decrease in width of the correlation function for modes with a higher transmission \cite{Shi2015}, as they are associated with narrow resonances \cite{Bliokh2008}. We observe an effect that seems exactly opposite, where it should be noted that we perform measurements in a 3D sample far from the localised regime. The difference between these regimes is intriguing and may lead to a new indicator of the approach of the localization transition.

In earlier work from our group, measurements on three dimensional diffuse samples have shown that the intensity of light focused through a medium, by optimizing intensity in a single spot, follows the speckle correlation function \cite{Beijnum2011}. However in that work an increase in total transmission was not measured. The key difference with this work is the performed optimization. While in Ref. \cite{Beijnum2011} the intensity in a single speckle spot was optimized and observed, here we optimize for, and observe, total transmission.

The spectral width of the transmission enhancement $\eta^{\mathrm{T}}_{\omega_0} (\Delta \omega)$ was found to be even broader than the correlation function. 
This is in line with simulations on 2D disordered waveguides in the diffusion regime \cite{Liew2014}. The broadening of the transmission enhancement with respect to the correlation function can be tentatively explained by the fact that it is only sensitive to decorrelation on the input of the sample. On the other hand the correlation function is also sensitive to dephasing between transmission channels on the output. The increase in width of the correlation function after optimization suggests that optimizing the total transmission leads to shorter transmission paths, decreasing the time light spends inside the sample. The relation to universal features of the delay time in diffusion is an intriguing aspect to be explored \cite{Pierrat2014}.

In conclusion, we have efficiently enhanced total transmission by coupling light to open transmission channels in a 3D strongly scattering sample. 
We have observed that enhancing the transmission by repeated phase conjugation leads to an increase in the frequency bandwidth of the field correlation function. In addition, we observe that the enhanced transmission persists over an even larger frequency bandwidth.
Our results show there is a subtle relationship
between transmission and transport delay time, which is largely unexplored theoretically and experimentally.

\section*{Acknowledgments}
We thank Jacopo Bertolotti, Ad Lagendijk, Pepijn W.H. Pinkse, Willem L. Vos and Hasan Y\i lmaz for discussions, Oluwafemi S. Ojambati for providing the sample and Cornelis A.M. Harteveld for technical support. This work is part of the research program of the Stichting voor Fundamenteel Onderzoek der Materie (FOM), which is supported by the Nederlandse Organisatie voor Wetenschappelijk Onderzoek (NWO). We acknowledge European Research Council (ERC) grant no. 279248 and NWO-Vici. 

\bibliographystyle{apsrev4-1}
 \bibliography{decorrelation}

\end{document}